\documentclass[twocolumn]{aastex62}
\usepackage{graphicx}
\usepackage{amsmath,amssymb}
\usepackage{slashed}
\usepackage{tcolorbox}
\usepackage{wasysym}
\usepackage{ulem}
\usepackage{color}
\usepackage{cancel}

\newcommand{\B}{{\scriptscriptstyle B}}
\newcommand{\jrs}[1]{\textsf{\color{blue}{ #1}}}

\definecolor{theo}{rgb}{1.0, 0.49, 0.0}

\graphicspath{{./}{figures/}}

\submitjournal{ApJ}

\begin{document}
	
	\title{Isovector Effects in Neutron Stars, Radii and the GW170817 Constraint}
	
	
	\author{T. F. Motta}
	\affil{CSSM and ARC Centre of Excellence for Particle Physics at the Terascale, \\ Department of Physics, University of Adelaide SA 5005 Australia}
	
	\author{A. M. Kalaitzis}
	\affil{CSSM and ARC Centre of Excellence for Particle Physics at the Terascale, \\ Department of Physics, University of Adelaide SA 5005 Australia}
	
	\author{S. Anti\'{c}}
	\affil{CSSM and ARC Centre of Excellence for Particle Physics at the Terascale, \\ Department of Physics, University of Adelaide SA 5005 Australia}

	\author{P. A. M. Guichon}
	\affil{IRFU-CEA, Universit\'{e} Paris-Saclay, F91191 Gif sur Yvette, France}
	
	\author{J. R. Stone}
	\affil{Department of Physics (Astro), University of Oxford, OX1 3RH United Kingdom}
	\affil{Department of Physics and Astronomy, University of Tennessee, TN 37996 USA}

	\author{A. W. Thomas}
	\affil{CSSM and ARC Centre of Excellence for Particle Physics at the Terascale, \\ Department of Physics, University of Adelaide SA 5005 Australia}

	\begin{abstract}
		An isovector-scalar meson is incorporated self-consistently into the quark-meson coupling description of nuclear matter and its most prominent effects on the structure of neutron stars are investigated. The recent measurement of GW170817 is used to constrain the strength of the isovector-scalar channel. With the imminent measurements of the neutron star radii in the NICER mission it is particularly notable that the inclusion of the isovector-scalar force has a significant impact. Indeed, the effect of this interaction on the neutron star radii and masses is larger than the uncertainties introduced by variations in the parameters of symmetric nuclear matter at saturation, namely the density, binding energy per nucleon and the symmetry energy. In addition, since the analysis of GW170817 has provided constraints on the binary tidal deformability of merging neutron stars, the predictions for this parameter within the quark-meson coupling model are explored, as well as the moment of inertia and the quadrupole moment of slowly rotating neutron stars.
		
	\end{abstract}


	\section{Introduction} \label{sec:intro}
	As the repositories of the densest nuclear matter in the Universe, neutron stars (NS) have long been a focus of nuclear theory studies (\cite{Glendenning:1997wn, Weber:1999qn, Haensel:2007yy, Lattimer:2012nd,Lattimer:2013zqa,Baym:2017whm}). The interest in these intriguing objects has intensified since the LIGO and Virgo observation of GW170817 (\cite{TheLIGOScientific:2017qsa}), an event identified as almost certainly the merger of two NS. The GW170817 analysis has already demonstrated the potential of gravitational wave (GW) observations to yield new information, such as the limits on NS tidal deformability, which has not hitherto been accessible.
	
	Before the observation by GW170817 but following the discovery of particularly heavy NS with masses around $2$ $M_{\odot}$ (\cite{Demorest2010,Antoniadis2013}), the astrophysics community was already making great efforts to model nuclear matter equations of state (EOS) capable of producing such heavy objects (see for example  (\cite{Oertel:2016bki}).
This was especially challenging since such heavy stars are expected to be so dense that, under the constraint 
of $\beta$-equilibrium, they must contain hyperons. The EOS including hyperons are typically much softer than those containing just nucleons, leading to maximum masses which are not compatible with observations. There are several approaches used to solve this issue, such as the inclusion of repulsive three-body forces involving hyperons or the introduction of a possible phase transition to deconfined quark matter at densities below the hyperon threshold (for more details see  Vida\~na {\em et al.}~(\cite{Vidana2015}) and references therein). 
	
	Within the quark-meson coupling model (QMC) (\cite{Guichon:1987jp,Guichon:1995ue,Saito:2005rv,review18})  NS with masses of order $1.9$ $M_{\odot}$ are produced with or without the inclusion of hyperons in the EOS (\cite{RikovskaStone:2006ta,Stone:2010jt,danielcarroll,DanielHybrid}). This remarkable feature, first published three years before the first heavy NS measurement (\cite{Demorest2010}), is a consequence of the fact that the self-consistent modification of the internal quark structure of hadrons in-medium naturally leads to repulsive three-body forces between all baryons  -- see Refs.~\cite{Guichon:2004xg,review18}. Furthermore, these forces involve {\em no} new parameters but are a direct consequence of the underlying quark structure, through the so-called scalar polarizabilities. Here, we extend the model presented in Ref.~\cite{RikovskaStone:2006ta} by adding  the exchange of the isovector-scalar meson $a_0(980)$, labelled $\delta$, in addition to the original isoscalar-scalar $\sigma$, isoscalar-vector $\omega$ and isovector-vector $\rho$ mesons and the pion, and investigate the effect of this modification on predictions of NS properties. We note that earlier calculations with the QMC model (without the Fock terms) have included such an interaction (see \cite{Santos}) 
to study low density instabilities in asymmetric nuclear matter, while 
\cite{Wang} used the QMC model with the isovector-scalar channel included in their Hartree-Fock calculation of properties of finite nuclei. This work extends, for the first time, the QMC model with the isovector-scalar exchange and the full Fock terms, to predict properties of high density matter in NS. 
	
	While NS masses can be determined reasonably accurately, the situation regarding their size is much less satisfactory. A number of techniques have been used to put broad limits on their radii, with values typically in the range $6 - 15 \, km$ (\cite{Ozel_Radii}). A  comprehensive recent update can be found in the work of Steiner {\em et al.} (\cite{Steiner:2017vmg}) and references therein. These limits do not constrain the EOS strongly enough and many different models of dense matter are still allowed. However, a great deal of anticipation surrounds the NICER experiment (\cite{Gendreau2016}), which aims to measure the radii of NS with an accuracy of order $5\%$ (\cite{Watts}) In anticipation of results from NICER as well as future observations of NS mergers, with GW170817 yielding its own constraint, our aim is to investigate to what extent the QMC model with the isovector-scalar interaction can provide reliable predictions for the radii of NS, with realistic error estimates. As the first positive sign we will show that the comparison of the QMC model predictions with the recent gravitational wave measurement already suggests an upper bound on the strength of the isovector-scalar coupling in this model. 

	The isovector-scalar exchange has been extensively explored in relativistic-mean-field models in the Hartree approximation ( see, for example, 
Refs.(\cite{Kubis:ew,Liu:2001iz, Menezes:2004vr,RocaMaza:2011qe,Singh:2014zza})). The isovector-isoscalar meson $\delta$ has also been included in the chiral mean field model (CMF)~\cite{Beckmann:2001bu}.  The Dirac-Brueckner-Hartree-Fock model (\cite{Sammarruca:2012vb}), including six non-strange bosons with masses below 1GeV, $\pi,\eta,\rho,\omega,\sigma$ and $\delta$, was also applied to nuclear matter and NS and the results were compared with the outcome of chiral effective field theory. The main conclusions of these investigations, relevant for nuclear matter, are that the effects of the isovector-scalar channel are almost negligible in symmetric nuclear matter but significant in matter with high asymmetry and thus revelant for modelling neutron stars. Predictions for the EoS,  the density dependence of the symmetry energy and its slope, the split of the proton and neutron effective mass and the composition of asymmetric matter, namely the proton fraction and the hyperon thresholds have been reported as a function of the isovector-scalar coupling strength, leading to estimates of its value.
	
	In section~\ref{sec:model} we describe the model used in this work. Fixed and variable parameters of the model are summarized in section \ref{sec:param}, followed by section \ref{sec:results} containing our main results. Conclusions can be found in section \ref{sec:conclusion}. 

	\section{Theoretical Model}
\label{sec:model}
	The QMC model (\cite{Guichon:1987jp,Saito:1994tq, Guichon:1995ue}) is based on the hypothesis that baryons, consisting of bags containing three confined, valence quarks, interact amongst themselves by the exchange of mesons that couple directly to the non-strange quarks. For vector mesons  the mean fields simply  shift the baryon energies, as in other relativistic models. On the other hand, a mean scalar field modifies the effective mass of the confined quarks, leading to a change in the valence quark wave functions. This in turn leads to the modification of the scalar field coupling to the hadron and hence, in order to find the effective baryon mass in-medium, one must solve the entire problem self-consistently (\cite{Guichon:1995ue,RikovskaStone:2006ta}). There is immense interest in looking for evidence of these changes in  baryon structure (\cite{Cloet,Cloet:2009qs}). However, for the present purpose, the essential result of the model is that the effective baryon mass in-medium no longer has a simple linear dependence on the mean scalar field. 
	
	Within the QMC model, the internal structure of a nucleon, or in general a baryon, is described by the MIT bag model (\cite{Bag}), within which the effective in-medium baryon mass with the flavour content $N_u, N_d, N_s$ is given as 
	\begin{equation}\label{efmass}
	M_B^{\star}=\frac{\Omega_uN_u+\Omega_dN_d+\Omega_sN_s}{R_B}-\frac{Z_0}{R_B}+\Delta E_{M}+\mathcal{B}V_B \,  .
	\end{equation}
Here $\Omega_q$ is the quark's lowest energy eigenvalue in the bag, taking into account the interaction with the mean scalar field, $Z_0$ is the so called zero point parameter that corrects the energy for gluon fluctuations and centre of mass effects, $\Delta E_{M}$ is the hyperfine colour interaction (\cite{DeGrand1975}), which is also modified by the applied scalar field (\cite{Guichon:2008zz}), while \jrs{$\mathcal{B}$} is the bag constant and $V_B$ the volume of the bag. 
	
	The effective mass can be written as
	\begin{equation}
	M_B^\star = M_B - g_{\sigma B}(\sigma, \delta)\sigma - g_{\delta B}(\sigma, \delta)\frac{\boldsymbol\tau\cdot\boldsymbol\delta}{2} \, ,
	\end{equation}
where the functions $g_{\sigma B}(\sigma, \delta)$ and $ g_{\delta B}(\sigma, \delta)$ are fitted to reproduce Eq.~(\ref{efmass}) for each baryon, as a function of the strength of the isoscalar ($\sigma$) and isovector ($\delta$) scalar fields. Thus, while in practice $g_\sigma, g_\delta, g_\omega$ and $ g_\rho$ (the couplings to the nucleon, $B=N$ in free space) are treated as parameters, these are directly related to the underlying coupling constants of the mesons to the quarks. These, in turn, allow calculation of the couplings to all of the hyperons with no new parameters.
	
	In practice it is sufficient to make an expansion up to terms quadratic in the meson field. So we write
	\begin{equation}
	M_\B^\star(\sigma,\delta)=M_\B - g_\sigma\sigma -t_{3\B} g_\delta  \delta + w_\B^\sigma\frac{\sigma^2}{2} +  w_\B^\delta\frac{\delta^2}{2} +  \lambda_\B{\sigma\delta} \, ,
	\end{equation}
where $t_\B$ and $t_{3\B}$ are respectively the isospin and the isospin projection of a baryon and $w_\B^{\sigma,\delta},\lambda_\B$ are weight parameters that we fit to reproduce the result of 
Eq.~(\ref{efmass}). 
        The expression for the energy density, including the full exchange Fock terms, derived in Ref.~\cite{RikovskaStone:2006ta},  which arise from single pion exchange, and the meson mean field equations can be found in the Supplementary Material. 
These equations are solved self-consistently, minimizing the energy density subject to the constraints of charge neutrality, $\beta$-equilibrium and baryon number conservation.

	\begin{figure}[b]
		\centering
		\includegraphics[width=\linewidth]{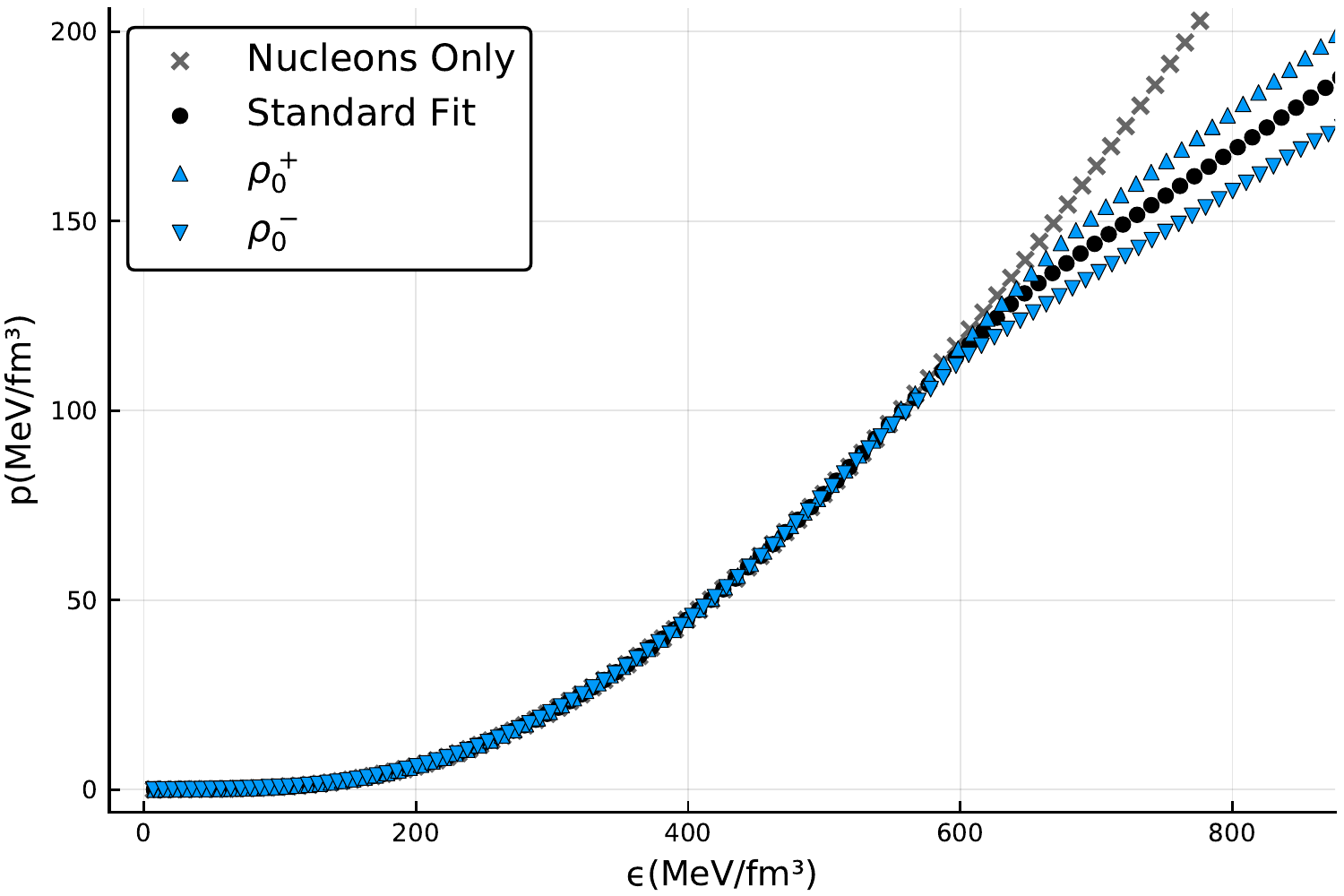}
		\caption{Illustration of the EOS for matter containing the full baryon octet for the different sets of parameters chosen here (see text and Table 1). A single  ``Nucleons Only'' EOS, for which the parameters reproduce the standard fit nuclear matter parameters, is also shown for comparison. Note that the up/down triangles refer directly to an increase/decrease of the saturation density.}
\label{fig:EOS}
	\end{figure}
%

	\section{Input parameters}\label{sec:param}
        The QMC model has two fixed parameters, the bag radius, set to $R_B$ = 0.8 fm and the mass of the scalar meson $\sigma$, $m_\sigma=700$ MeV, which is not well known experimentally. The masses of  the pion, $\omega$, $\rho$  and $\delta$ and  of the  baryon octet were taken from the experiment. The bag constant, strange quark mass and the colour interaction strength, $\alpha_c$ are fitted within the model to reproduce the bare mass of the free nucleon and the $\Lambda$-hyperon, as well as to fulfil the stability condition $\partial_{R_B}M^\star_B=0$.

The variable parameters, the coupling constants  $G_{\phi}$ for the mesons, $\phi = (\sigma, \omega, \rho, \delta)$, to the nucleon are defined as $G_{\phi} =g_{\phi}^2/m_{\phi}^2$. These are obtained by fitting to the empirical properties of symmetric nuclear matter, the saturation density, $\rho_0$, the binding energy per particle, $\mathcal{E}$,  and the symmetry energy, $S$, at saturation. The values $\rho_0=0.16 \rm{ fm}^{-3}$, $\mathcal{E}=-15.8 \rm{MeV}$ and $S=30 \rm{MeV} $ were chosen as `standard' in this work. In order to probe the effects of a change in nuclear matter parameters on the nuclear matter EOS and, consequently, the NS gravitational mass and radius, six combinations of the parameters, deviating from the standard values, were constructed to form additional EOSs. 	Fits labeled $\rho_0^+$,$\rho_0^-$ deviate by $\pm0.01fm^{-3}$ from the standard value of the saturation density, as detailed in Table~\ref{tab:param}. We also allow variations of the symmetry energy and binding energy of $\pm2$MeV around the standard fit value. For clarity, we refrain from showing the results for fits other than the standard and $\rho_0^\pm$ as all results for other parameter variations were found to lie between the results for $\rho_0^\pm$ fits.

The range within which each parameter is varied defines the uncertainty band in $\rho_0$, $S$ and $\mathcal{E}$ around the Standard EOS. Each set has been used to determine the three couplings  $G_{\sigma}$,  $G_{\omega}$ and  $G_{\rho}$. The coupling $G_{\delta}$ was kept constant equal to 3.0 fm$^2$ during this procedure. However, when examining the isovector effects,  $G_{\delta}$ was also varied about the central value of $3.0$ fm$^2$ (taken from the one-boson-exchange potential of \cite{Haidenbauer}). The couplings $G_{\sigma}$,  $G_{\omega}$ and  $G_{\rho}$ were also obtained for two other choices of the $\delta$-nucleon coupling, $G_{\delta}$=0 and 6.0 fm$^2$ (denoted $2 G_{\delta}$) and these are shown in the Supplementary Material.   
	{\begin{figure}[t!]
			\begin{center}
				\includegraphics[width=\linewidth]{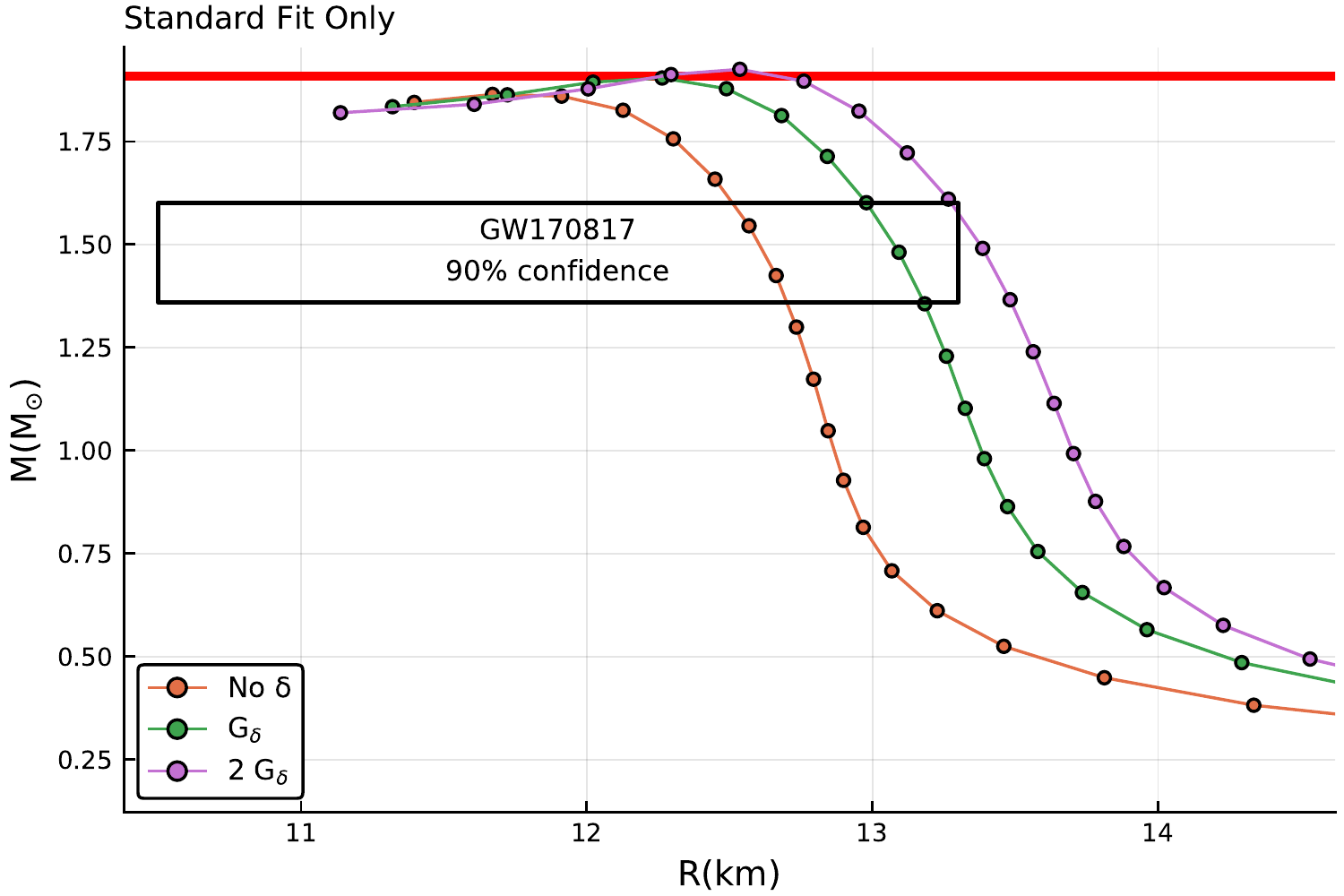}
				\caption{Mass vs Radius diagrams for different values of the $\delta$ coupling constant, calculated using only the Standard Fit parameters for nuclear matter  (see Table \ref{tab:param}). All curves include the effect of hyperons at sufficiently large density. The box shows the region preferred by the joint LIGO-Virgo analysis~\cite{Abbott:2018exr}, while the red band indicates the currently preferred mass of PSR J1614−223~\cite{Arzoumanian}.}
				\label{fig:variastov}
			\end{center}
	\end{figure}}
	{\begin{figure}[t!]
			\begin{center}
				\includegraphics[width=\linewidth]{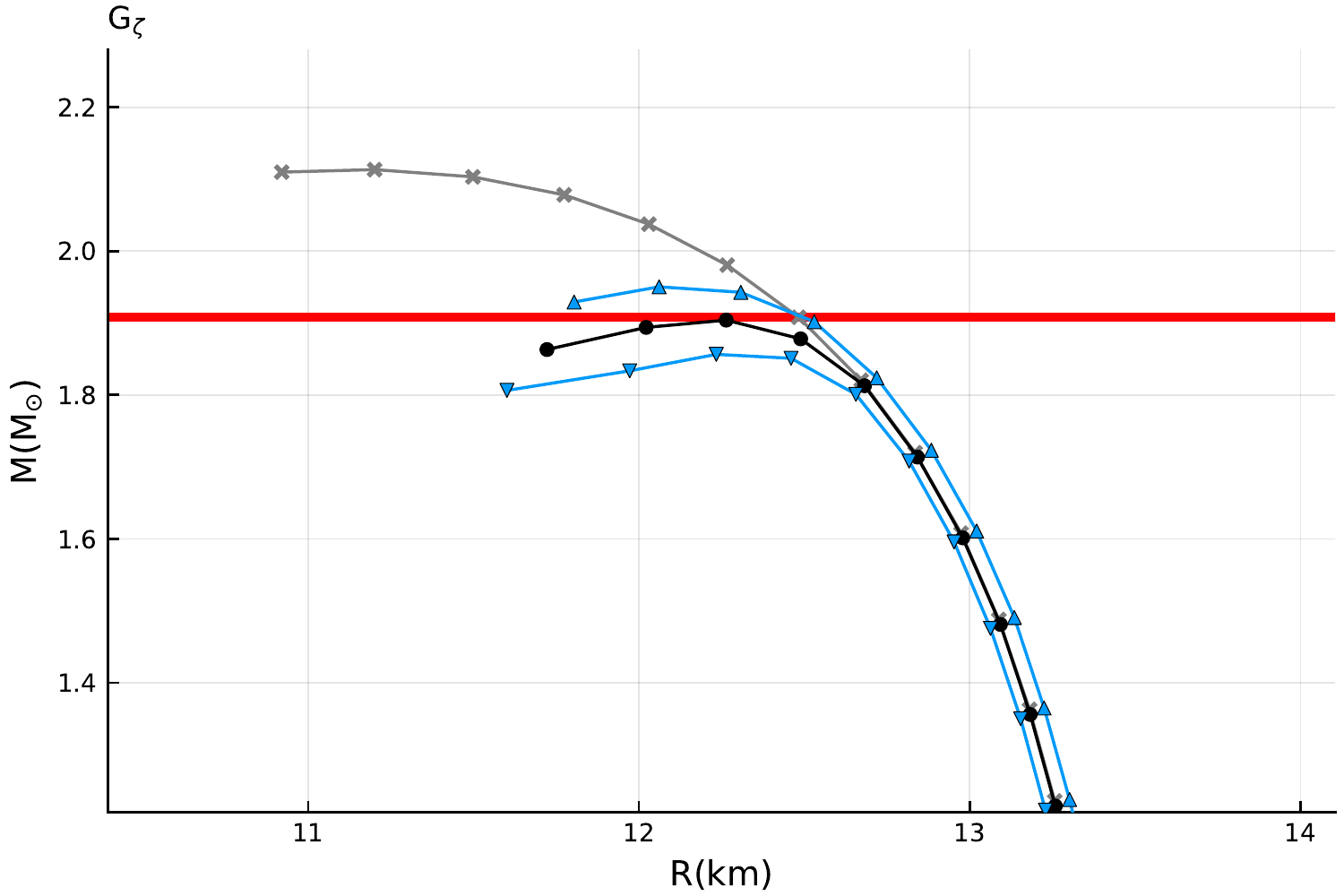}
				\caption{Mass vs Radius diagrams for the variations of nuclear matter parameters shown in Table~\ref{tab:param} for the standard $\delta$ coupling strength. The symbols are as defined in Fig.~1.}
				\label{fig:TOVs}
			\end{center}
	\end{figure}}
	\begin{table*}[t]
		\centering
		\bgroup
		\def\arraystretch{1.2}
		\begin{tabular}{l c c c c c c c c c c c c} 
			\hline
			\hline
			EOS 				& & \multicolumn{4}{l}{NM parameters} & \multicolumn{7}{l}{QMC couplings}   \\  
			& &$\rho_0$ & $S$   & $\mathcal{E}$ & & $G_{\omega}$ & $G_{\rho}$ &$G_{\sigma}$ &$G_{\delta}$ & &$L_0$ & $K_0$ \\
			& &[fm$^{-3}$] & [MeV] & [MeV] & & [fm$^{2}$] & [fm$^{2}$] & [fm$^{2}$] & [fm$^{2}$] & &[MeV] &[MeV] \\
			\hline
			\textbf{Std. Fit}   &  & $0.16$     & $30$ & $-15.8$ & &  6.39& 4.27& 10.0& 3.0
			& &63.7& 282   \\
			{$\boldsymbol \rho_0^+$}   &  & $0.17$     & $30$ & $-15.8$ & & 5.93& 3.81& 9.46& 3.0
			& &62.6& 284      \\
			{$\boldsymbol \rho_0^-$}   &  & $0.15$     & $30$ & $-15.8$ & &  6.88& 4.59& 10.64& 3.0 & &
			62.4& 280    \\
			\hline
			\hline        
		\end{tabular}
		\caption{Nuclear matter parameters and coupling values for different EOSs determined by the values of $\rho_0$, $S$ and $\mathcal{E}$. The value of $K_0$ does not change significantly for different values of $G_\delta$ , while the slope, $L_0$, for the cases with $G_\delta=$ 0 and 6 fm$^2$ we have, respectively, a decrease and an increase of 10 MeV.
\label{tab:param}}
		\egroup
	\end{table*}
	
\section{Results}\label{sec:results}
	
	By varying the nuclear matter parameters over the specified ranges we can define an uncertainty band regarding the variation in $\rho_0$, $S$ and $\mathcal{E}$ parameters around the Standard EOS, as given in Fig.~\ref{fig:EOS}. In order to calculate the M-R curves shown in Figs.~\ref{fig:variastov} and \ref{fig:TOVs}, the solution of the TOV equations using the QMC model to describe the NS core is matched to the low-density EOS that accounts for the NS crust. In order to describe the crust we have used the  EoS developed by Hempel {\em et al.}~(\cite{Hempel:2011mk,Hempel:2009mc}), which was also recently used by Marques {\em et al.}~(\cite{DD2Y}) to augment their EoS of a NS core based on the relativistic mean field model with the density dependent interaction DD2Y, including the full baryon octet. This choice is similar to that of the QMC model and we regard it as sufficiently realistic for the purpose of this work. In particular, for stars of mass above 1.4 M$_{\astrosun}$ the crust occupies only the outer 20\% or less of the radial profile of the star.

From Fig.~\ref{fig:variastov} we see the fairly dramatic change in radius for a typical neutron star 
as $G_\delta$ is varied. That this is considerably larger than the variation associated with the choice of nuclear matter parameters may be seen in Fig.~\ref{fig:TOVs}. 
	\begin{figure}[t!]
		\centering
		\begin{minipage}{.45\textwidth}
			\includegraphics[width=1\textwidth]{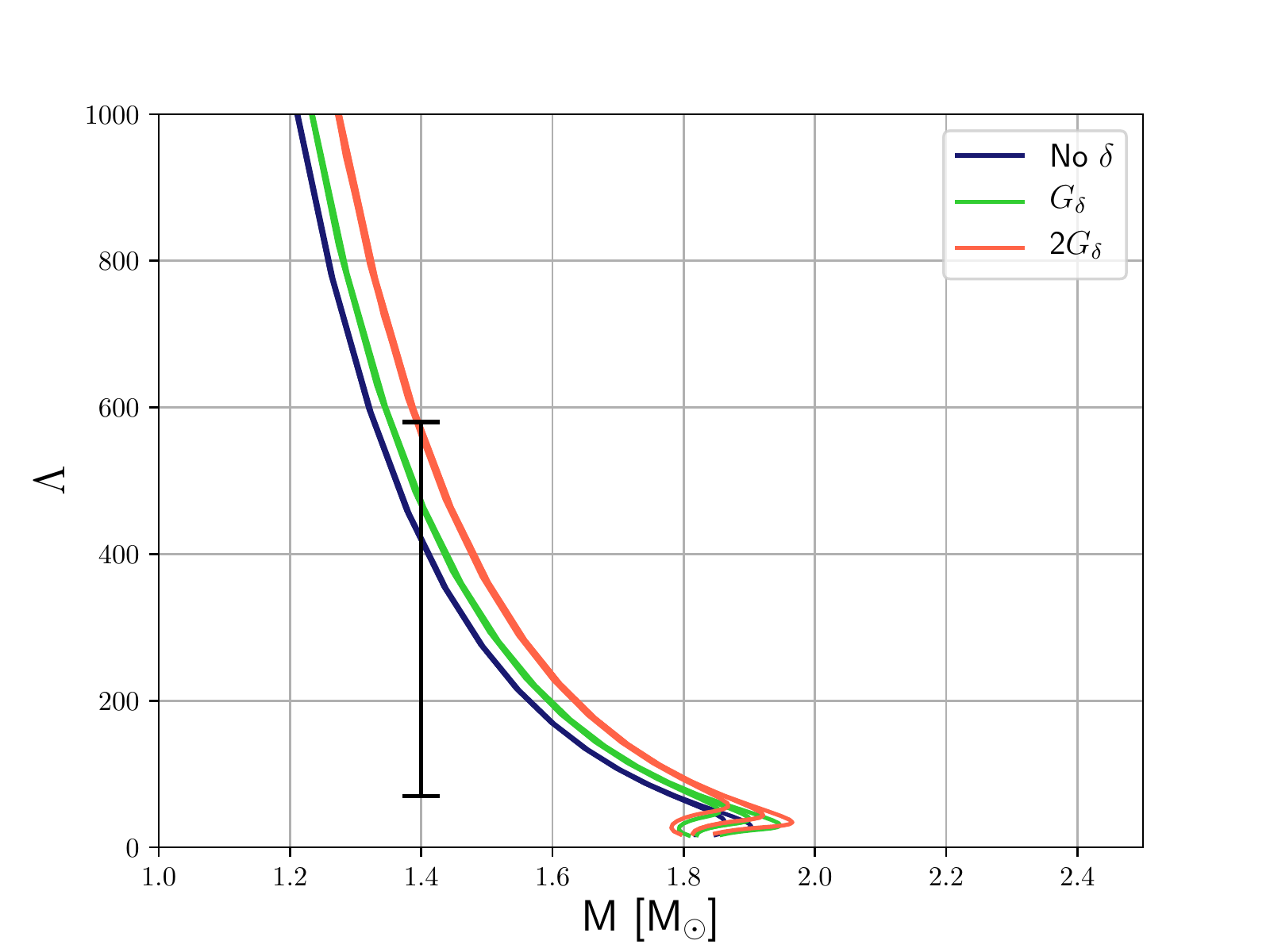}
			\caption{Tidal deformability as a function of the mass. The width of the curves illustrates the relatively small dependence on the choice of nuclear matter parameters. 
The bar at $M=$1.4M$_\odot$ shows the constraints derived in Ref.~\cite{Abbott:2018exr}.}\label{fig:TD}
		\end{minipage}\hspace{0.01\textwidth}
		\begin{minipage}{.45\textwidth}\label{Lambda}
			\includegraphics[width=1\textwidth]{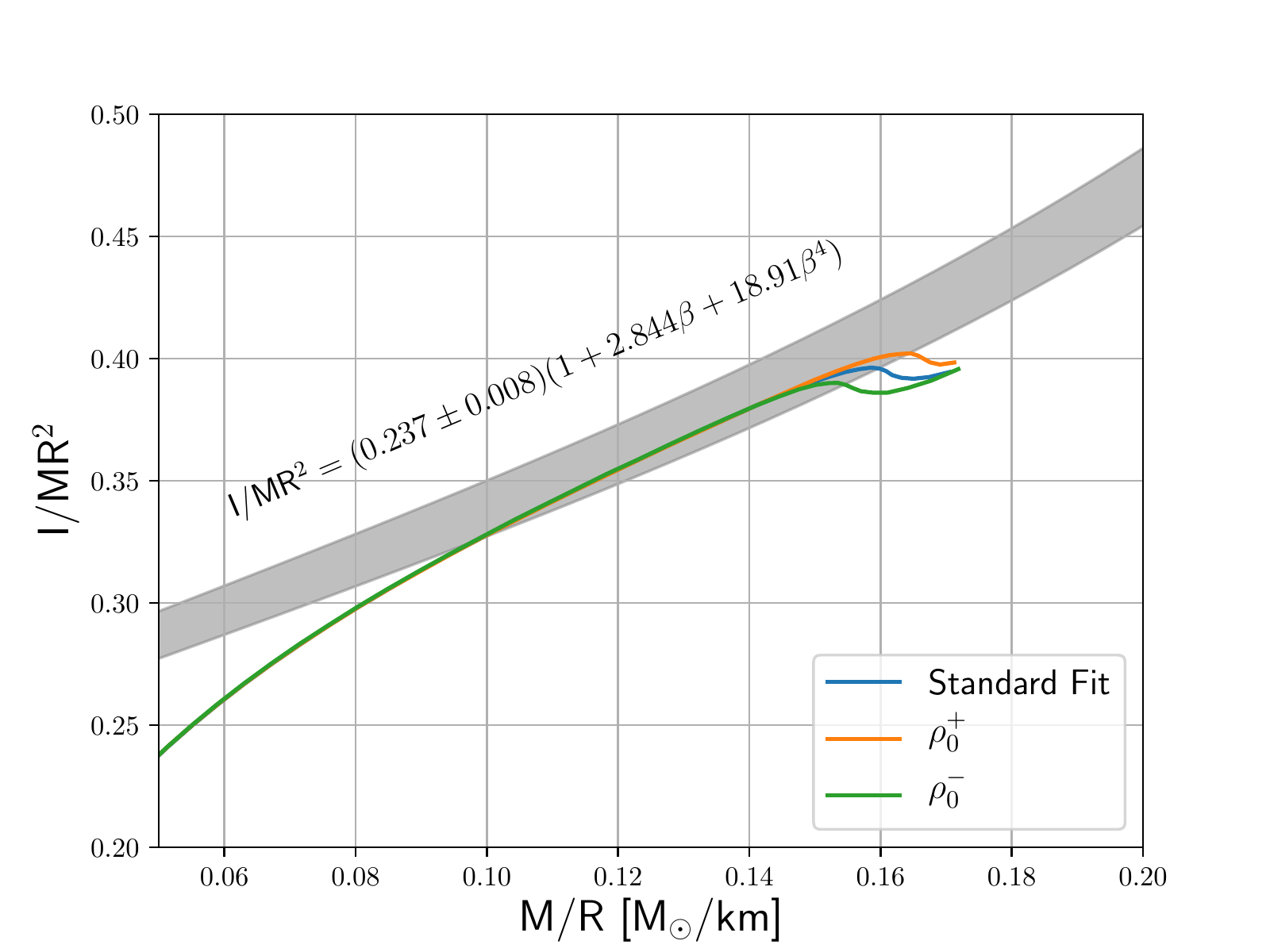}
			\caption{Moment of inertia of the star versus the ratio $M/R$, compared with the region preferred in the analysis of Ref.~\cite{Zhao:2018nyf}.}
\label{MoI}
		\end{minipage}
	\end{figure}
\begin{figure}[b]
	\centering
	\includegraphics[width=\linewidth]{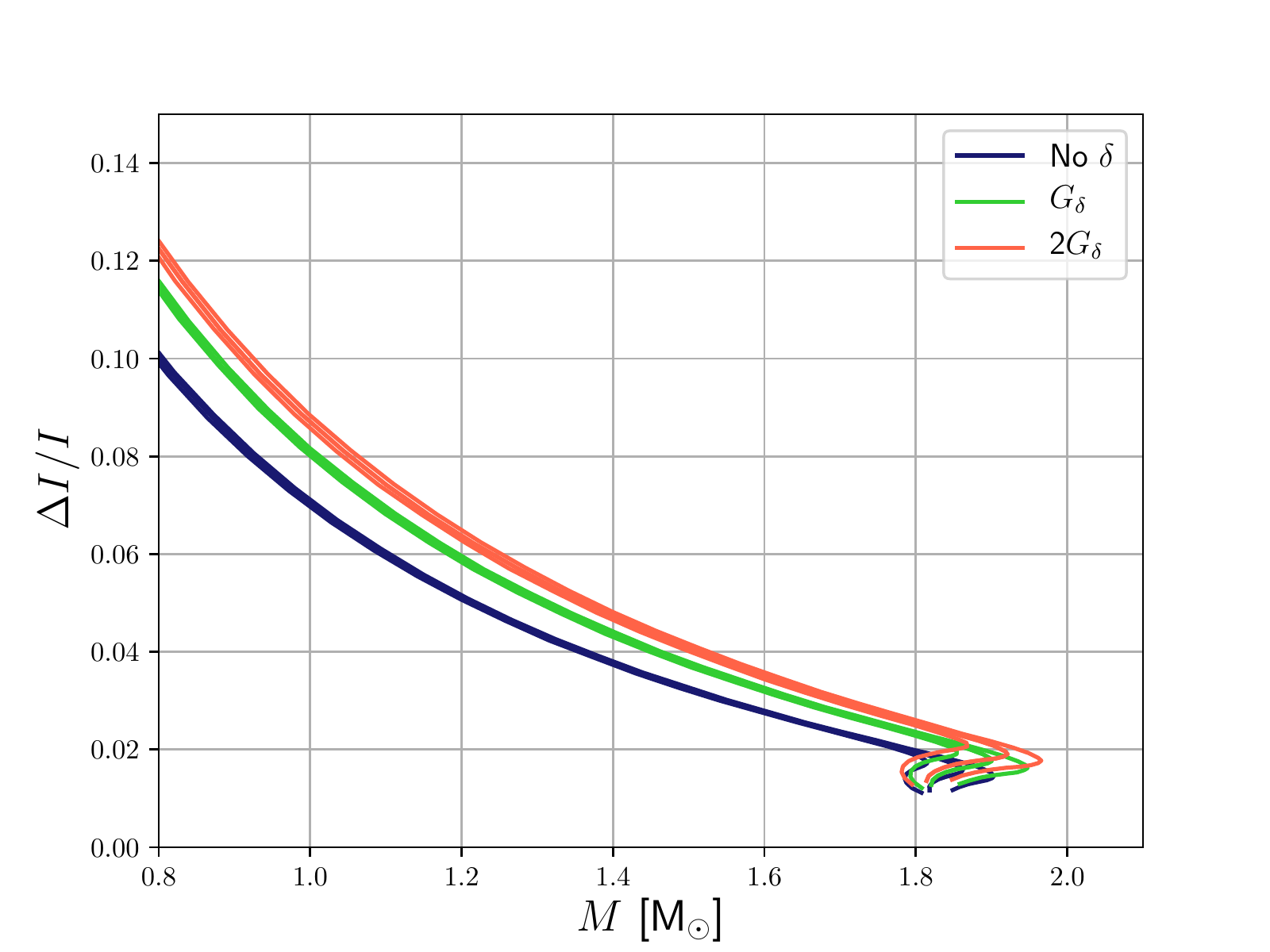}
	\caption{Fraction of the total contribution to the moment of inertia coming from the crust. The colours indicate the value of $G_\delta$, while the width of each curve indicates the effect of the variation of nuclear matter parameters.}
	\label{fig:CrustMoI}
\end{figure}
	\begin{figure}[t!]
		\begin{center}
			\includegraphics[width=\linewidth]{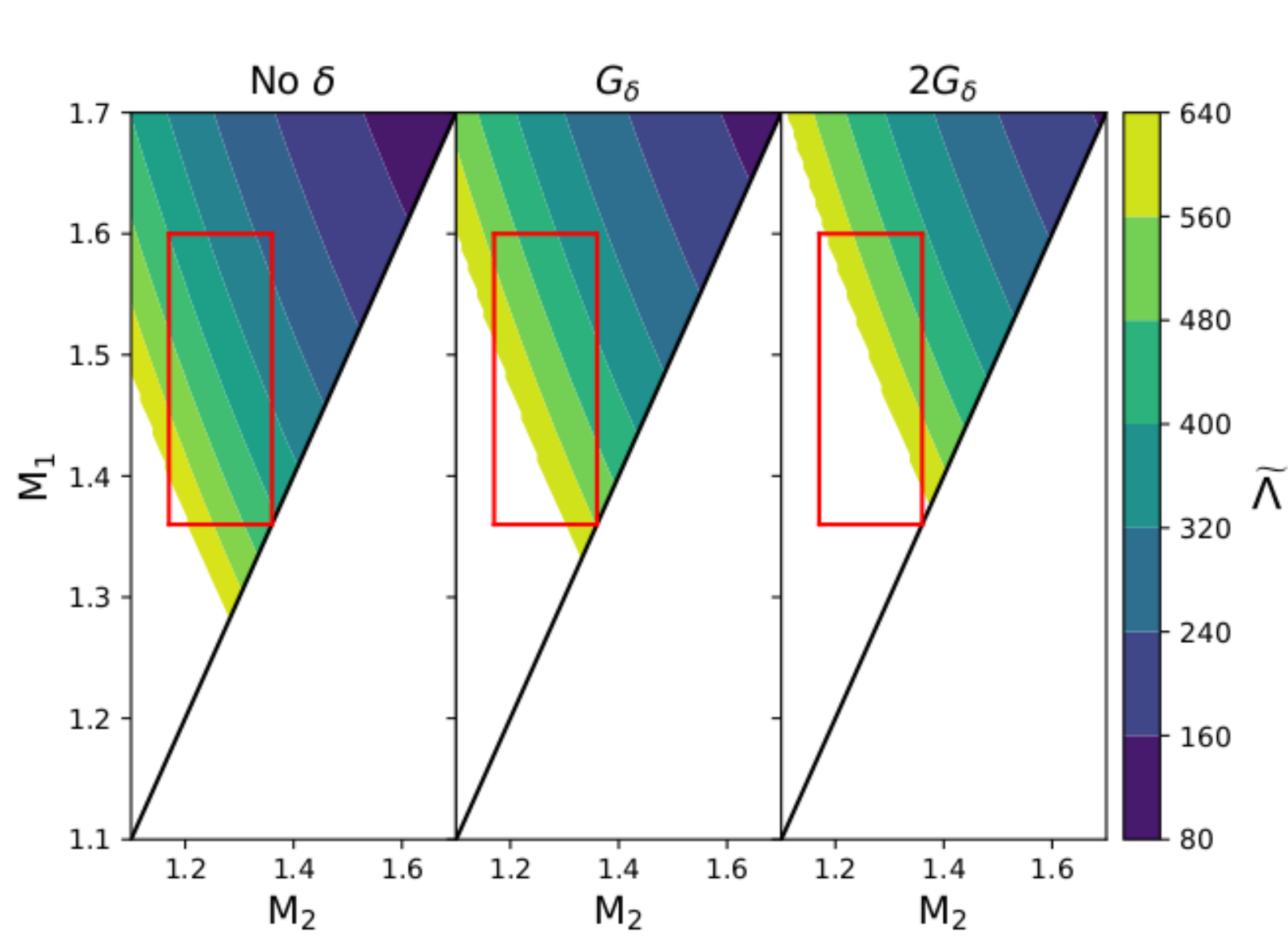}
			\caption{Binary tidal deformability $\tilde{\Lambda}$, with the white area above the line $M_1 =M_2$ being excluded by the constraint $80<\tilde{\Lambda}<640$~(\cite{Steiner,De:2018uhw})}.
			\label{fig:lambtilda}
		\end{center}
	\end{figure}

	Next we examine the effect that the introduction of the $\delta$-meson has on NS structure. The gravitational mass of the NS, obtained by solving the Tolman-Oppenheimer-Volkoff (TOV) equations (\cite{Oppenheimer:1939ne}) using the standard EoS, is shown in  Fig.~\ref{fig:variastov} as a function of the corresponding radius for three different values of $G_{\delta}$ . We observe that the $\delta$ coupling strength has a sizable effect on the radius, while changing the maximum mass by less than $0.15$M$_{\astrosun}$. 
	
	It is especially interesting to observe that, for the standard choice of nuclear matter parameters (Standard Fit in Table~\ref{tab:param}), the presence of the $\delta$ meson shifts the M-R curve toward larger radii.  By centering our search around $G_\delta=3.0$ fm$^2$, one sees clearly that the value of this coupling produces transverse shifts of the whole M-R diagram that exceed the spread arising from variations in the nuclear matter parameters, which is shown in Fig.~\ref{fig:TOVs}. For NS with gravitational masses around $1.5$ M$_{\astrosun}$, the inclusion of the isovector channel results in radius changes of the order of $0.5$ km. The recent LIGO - Virgo joint analysis of the GW170817 data (\cite{Abbott:2018exr}) constrains the NS radius (at 90\% confidence level) to the black box given in 
Fig.~\ref{fig:variastov}. This upper limit on the radius is consistent with other recent work~\cite{Raithel:2018ncd,Most:2018hfd,Annala:2017llu} and suggests  that within the QMC model the 
strength of the isovector-scalar sector is most likely to be $G_\delta \lessapprox 6.0$fm$^2$.
	
	Next we demonstrate the effect of the isovector scalar channel on two macroscopic properties of NS, the moment of inertia, $I$, and the tidal deformability, $\Lambda$. As we see in Fig.~\ref{fig:TD}, the variation of this parameter with the strength of the $\delta$ coupling is relatively weak. Furthermore, we have verified the universality of the relation between the moment of inertia, the Love number and the quadrupole moment~\cite{Hartle:1967he,Hartle:1968si}, supporting the proposal that this is independent of the EoS~\cite{Yagi:2013awa}. Given that there is also interest in the moment of inertia calculated within the model, we show in Fig.~\ref{MoI} the results of the calculation using the slow-rotation Hartle-Thorne approximation~\cite{Hartle:1967he} and an explicit comparison with the work carried out in Ref.~\cite{Zhao:2018nyf}. It is clear that the moment of inertia calculated in the present model is consistent with the constraint region proposed there. 

	Measurement of the moment of inertia of highly relativistic double pulsar system such as PSR J0737-3039 may be within reach after a few years of observation and yield a result with about 10\% accuracy~\cite{Steiner}. Given that the masses of both stars in that system are already accurately determined by observations, a measurement of the moment of inertia of even one neutron star would enable accurate estimates of the radius of the star and the pressure of matter in the vicinity of 1 to 2 times the nuclear saturation density. This,  in turn, would provide strong constraints on the equation of state of neutron stars and the physics of their interiors. Our calculations show little sensitivity to the variation in nuclear matter parameters, except for the region of compactness between 0.15 and 0.16 [M$_\odot$/km] where they differ slightly from the results reported in 
Ref.~(\cite{Zhao:2018nyf}). In the compactness region between 0.10 - 0.15 they are in agreement of the lower bound of the Zhao and Lattimer analysis~(\cite{Zhao:2018nyf}). We illustrate the contribution of the low density region to the moment of inertia in Fig.~\ref{fig:CrustMoI}, showing that, for stars with masses greater than $1$M$_\odot$, the low density region contributes less than $10\%$ of the total, going below $5\%$ for masses $\geq1.4$M$_\odot$ for all fits and all values of the $\delta$ coupling.
	
	Using the signal from GW170817, a restriction was placed on the binary tidal deformability, ${\tilde\Lambda}$, namely that it should lie inside the window $80<\tilde\Lambda<640$~(\cite{Zhao:2018nyf}), while the masses of the stars in the binary system were in the window $(1.36<m_1<1.6)\times(1.17<m_2<1.36)$. This constraint is indicated by the red box in Fig.~\ref{fig:lambtilda}, where we compare the present model calculations for the different strengths of the $\delta$ meson coupling $G_{\delta}$. Calculations of such quantities were performed following the work of Refs.~\cite{Hartle:1968si,Zhao:2018nyf,Postnikov}. Clearly the QMC model results overlap the constraint region and constrain the NS masses and the values for $\tilde{\Lambda}$ even further. The measurement does not rule out the model with double the delta coupling $2G_\delta$ but, as we can see in Fig.~\ref{fig:variastov}, it does tend to give values for the radius which are in some tension with the preferred region. 
	
\section{Conclusion}\label{sec:conclusion}
The QMC model, where the quark structure of the baryons adjusts self-consistently to the mean scalar fields generated in medium, has been extended to include the isovector-scalar meson, $\delta$. The self-consistent change in structure leads to repulsive three-body forces between all baryons (nucleons {\em and} hyperons), without additional parameters. Because of these forces the model still (c.f. Ref.~\cite{RikovskaStone:2006ta}) yields maximum neutron star masses of the order $\gtrapprox 1.9$ M$_{\astrosun}$. The major effect of the $\delta$ meson on NS properties is to increase their radii (c.f. Fig.~\ref{fig:variastov}), with the effect being considerably larger 
than the effect of varying the saturation properties of symmetric nuclear matter (c.f. Fig.~\ref{fig:TOVs}).

We have studied the moment of inertia ($I$) and tidal deformability ($\Lambda$) of NS  over a wide range of masses. In the case of the moment of inertia our results are consistent with the constraint suggested by Zhao and Lattimer ~(\cite{Zhao:2018nyf}) on the variation of $I/MR^2$ versus $M/R$. Our results also support the universal relation between $I$, the Love number and $\Lambda$, suggesting that it is indeed independent of the EoS used.
	
Following the neutron star merger observed recently by LIGO and Virgo, we also explored the dependence of the binary tidal deformability, $\tilde{\Lambda}$, on the parameters of the model as well as the masses of the stars involved. The results of our analysis tend to favour the larger radius end of the constrained region in Fig.~\ref{fig:variastov}, while being consistent with the binary tidal deformability constraints reported in Refs.~\cite{Abbott:2018exr,Lattimer:2004nj}.

\section*{Acknowledgements}
This work was also supported in part by the University of Adelaide and by the Australian
Research Council through the ARC Centre of Excellence for Particle Physics at the Terascale
(CE110001104) and Discovery Projects DP150103101 and DP180100497.	
	
	\bibliography{bibliography}{}
	\bibliographystyle{abbrvnat}
	\setcitestyle{authoryear,open={((},close={))}}

\end{document}